\documentclass[twocolumn,10 pt,superscriptaddress,amsmath,amssymb,showpacs]{revtex4}
\usepackage{graphicx}
\begin{document}

\title{Metastable states of a flux line lattice studied by transport and Small Angle Neutron Scattering}
\author{A. Pautrat}
\email[corresponding author: ]{alain.pautrat@ismra.fr} \author{J.Scola}\author{Ch.Simon}
\affiliation{CRISMAT/ENSI-Caen, UMR 6508 du CNRS,6 Bd Marechal Juin, 14050 Caen, France.}
\author{P. Mathieu}
\affiliation{Laboratoire Pierre Aigrain de l'Ecole Normale Sup\'erieure, UMR 8551 du
CNRS, associ\'ee aux universit\'es Paris 6 et 7, 75231 Paris Cedex5, France.}
\author{A.\ Br\^{u}let}
\affiliation{Laboratoire L\'{e}on Brillouin, \\ CEN Saclay, 91191
Gif/Yvette, France.}
\author{C. Goupil}
\affiliation{CRISMAT/ENSI-Caen, UMR 6508 du CNRS, Caen, France.}
\author{M. J. Higgins}
\affiliation{NEC Research Institute, 4 Independence Way, Princeton, New Jersey 08540.}
\author{S. Bhattacharya}
\affiliation{Tata Institute of Fundamental Research, Homi Bhabha Road, Mumbai 400005,
India.}
\begin{abstract}

Flux Lines Lattice (FLL) states have been studied using transport measurements and Small
Angle Neutron Scattering in low T$_c$ materials. In Pb-In, the bulk dislocations in the
FLL do not influence the transport properties. In Fe doped NbSe$_{2}$, transport
properties can differ after a Field Cooling (FC) or a Zero Field Cooling (ZFC) procedure,
as previously reported. The ZFC FLL is found ordered with narrow Bragg Peaks and is
linked to a linear V(I) curve and to a superficial critical current. The FC FLL pattern
exhibits two Bragg peaks and the corresponding V(I) curve shows a S-shape. This can be
explained by the coexistence of two ordered FLL slightly tilted from the applied field
direction by different superficial currents. These currents are wiped out when the
transport current is increased.
\end{abstract}


\newpage
\maketitle

Flux Line Lattice (FLL) order and its relationship with the pinning and dynamical
properties provides an excellent model system \cite{giam}. The competition between the
FLL elastic properties and the quenched or thermally driven disorder can lead to
different vortex matter states. To some extent, this implies also changes in the
transport properties. A good example is the peak effect observed in several type II
superconductors, a sudden increase of the critical current close to the
superconducting-normal transition, which has been for a long time considered as a proof
of a disordering transition. The basic idea was from Pippard \cite{pippard} who noted
that the pinning threshold for dilute bulk pinning goes quadratically
 to zero near the second critical field. Since this is faster than the elastic interaction between vortices, bulk pinning centers can become more effective on a less rigid lattice at
high field and that can lead to a peak effect. Larkin-Ovchinikov collective pinning model
\cite{larkin} and the associated scaling arguments gave a more precise theoretical
foundation to the link that is usually made between the loss of FLL order and the high
critical current. It follows that a high critical current is $\textit{a priori}$
associated to a disordered FLL. Nevertheless, these arguments suggest that the peak
effect should occur more often than it is actually found. Numerous experiments have also
shown that the link between the FLL order and the critical current is far from being
direct, in contradiction with the previous assumptions. Thorel's neutron scattering
pioneering experiments have first shown that the Flux Lines Lattice (FLL) quality can be
modified without changing the critical current \cite{thorel}. One possible explanation is
that the relevant length scale associated with the critical current is not the length
scale of topological order \cite{giamar}. Another possibility, already proposed by
different authors \cite{ms,archie,degennes}, is that the order of the FLL in the bulk is
not the most important parameter for understanding the transport properties of the FLL.

In addition to the peak effect, very peculiar transport properties can be observed. In
particular, whereas Voltage(Current) (V(I)) characteristics are usually reversible and do
not depend on the way from which the FLL is prepared, hysteretic V(I) curves are observed
in $NbSe_2$, when the FLL is formed after a Field Cooling (FC)
\cite{Kinnon,Bhattacharya}. A model has recently emerged, and is supported by different
experiments \cite{capacitor}. The key ingredients of this model are a supercooling  of a
high temperature/high critical current state into a low temperature/low critical current
state, and an annealing effect over surface barrier. In order to explain the high
critical current phase, a strongly disordered FLL is involved. This amorphous or
liquid-like state is in favor of a genuine phase transition when crossing the peak
effect. Nevertheless, very little is known about the real structure of these phases.
There is even contradictory and puzzling results. Indeed, recent decoration experiments
have shown that no disordering transition can be evidence in the peak effect region of
pure or Fe doped $NbSe_{2}$ samples \cite{menghini}. The high critical current FLL state
remains unexplained. Nevertheless, it should be specified that such experiments give the
position of field lines only when they protrude from the sample surface. It can not be
excluded that this distribution differs in the bulk.

 Small Angle Neutron Scattering appears as a complementary technique, since the
order of the FLL can be investigated \textit{in the bulk} of the material. It is also
possible to measure in real time V(I) curves together with the FLL diffraction patterns
and hence to investigate the relationship between the transport current properties and
the FLL structure. The aim of the following experiments is to use SANS in order to
compare different FLL states with respect to their dynamical properties. After discussing
a more conventional case, we will focus on the FLL states close to the peak effect in
NbSe$_{2}$.

The SANS experiments were performed in the Laboratoire Leon Brillouin (Saclay, France).
Large single crystals of Fe doped 2H-NbSe$_{2}$ (200 ppm of Fe, size $8 \times 6 \times
0.5 mm^{3}$, $T_{c}$ = 5.5 $\pm$ 0.15 K measured by specific heat) and of polycrystalline
Pb-In (10.5 $\%$ of In by weight, size $30 \times 5.5 \times 0.5 mm^{3}$, $T_{c}$= 7.05
$\pm$ 0.05 K) were used. The magnetic field was applied parallel to both the c-axis of
the crystal and the incident neutron beam. The scattered neutrons ($\lambda_{n}$ = 10
$\AA$, $\triangle \lambda_{n}$ / $\lambda_{n}$ $\approx 10 \%$) were detected by a 2D
detector located at a distance of 6.875 m. In the following, $\omega$ will refer to a
rotation around the vertical axis, and $\phi$ to a rotation around the horizontal axis
(Fig. 1). Superconducting leads were attached using Indium solder pressed between copper
slabs. At the working temperature of 2 K (in superfluid liquid He), this allows an
injection of about 10 A without any overheating. We present here results obtained for a
magnetic field of 0.4 T and 0.2 T.

Before describing the results obtained in Fe doped 2H-NbSe$_{2}$, it can be helpful to
present a more conventional case, where no peak effect and no hysteretic V(I) curves are
observed. In Pb-In samples, the dynamical properties of the FLL are well documented. For
a state of the FLL phase diagram (B and T fixed), only one critical current is measured.
The V(I) curves are found reversible when they are cycled. It does not necessary mean
that the FLL is always the same and that no metastable states are present, but it rather
means that the structure of these possible states does not play a significant role for
pinning and transport properties. For example, we have reproduced an experiment described
in \cite{houston}, but focusing on the details of the V(I) curves. The goal is to follow
the evolution of a moving FLL in a polycrystal of Pb-In (Fig. 2). The 2D pictures consist
in 12 acquisitions taken at a particular ($\omega_{i}$, $\phi_{i}$) angular position, in
order to fulfill at the best the Bragg conditions. Without any applied current, we
observe a powder-like diffraction pattern meaning that the FLL is highly dislocated and
that its orientational order is strongly degraded. This is due to the interaction between
the randomly oriented crystal axis of the sample and the FLL unit cells. The FLL is
ordered within the grains of the Pb-In slab but FLL dislocations are present in order to
accommodate the shear strains. The size of the crystal grains is optically evaluated to
be about 100 $\mu m^{2}$, which indeed corresponds to large ordered domains (thousands of
vortices). On the other hand, at the scale of the sample, this corresponds also to
thousands of ordered domains. This is enough to observe a ring of scattering due to the
poor azimuthal resolution of this technique. When increasing the current above the
critical current, the FLL order becomes strongly improved. The dislocations are expelled
when the flux lines are moving in the sample and long range orientational order is
established. The corresponding ordered state can be frozen (experimentally, we turned off
the current during the FLL flow). On the contrary, dislocations are appearing again when
the current is slowly decreased, which means that they can be considered as equilibrium
features. It is thus possible to stabilize a FLL with crystalline order or a FLL with
many dislocations and to analyze the V(I) curve corresponding of each FLL state. We
measured in both cases the same critical current and even exactly the same V(I) curve
(Fig. 2). These results show that the presence of large scale bulk dislocations in the
FLL governs orientational order but may have no effect on the critical current or on the
main dynamical properties. One should also refer to Thorel $\textit{et al}$ \cite{thorel}
who observed  different qualities of the vortex crystal that do not affect the critical
current, even in monocrystalline slabs of Niobium. In summary of this part, large scale
bulk dislocations are found to have no link with the critical current and with the
dynamical properties of the FLL. They can not be $\textit{a priori}$ involved to
interpret a high critical current state of the FLL.

We performed the same kind of experiments in crystals of $NbSe_{2}$, in order to compare
the different states of the FLL which could be responsible for the anomalous transport
properties. We have first observed the simplest case: the FLL after a Zero Field Cooling
(ZFC). Fig. 3a shows the SANS pattern we have obtained. This corresponds to an ordered
hexagonal FLL. The scattering wave vector $Q=0.00953 \pm 0.00050 \AA^{-1}$ is in good
agreement with the theoretical one, $Q_0=0.00938 \AA^{-1}$ calculated for the regular
crystal of flux lines. The alignment of the FLL is along both the a-axis and the lateral
faces of the crystal. No difference is observed on the 2D pictures for different FLL
velocities $V_L$ ($V_{L}=E/B$, with E the electric field and B the magnetic field). The
orientational order is thus preserved, confirming that, as it was observed in other type
II superconductors \cite{spinecho}, the FLL is well ordered and moves as a whole during
the flux flow. We also performed $\omega$ rocking curves. Small widths are obtained by
analyzing the peaks with Lorentzian fits (Fig. 3b). We obtain $\Delta\omega$ $(FWHM)$ =
0.232 $\pm$ 0.020 deg for the FLL after ZFC and without external current applied. This is
close to the experimental resolution given by the angular divergence of the beam (0.150
deg). If we increase the transport current, but staying below the critical current value,
absolutely no change is observed. When the applied current is higher than the critical
current I$_c$ = 2.5 A, a slight increase of the RC width $\triangle \omega$ is observed.
The reason is that the transport current $I$ imposed by the external generator has to
fulfill the Maxwell-Ampere equation. As the moving Bragg planes are observed to be
strictly translationally-invariant \cite{spinecho}, one can neglect the in-plane field
gradient and the Maxwell-Ampere equation reduces to $\mu_{o} J_y =\partial B_x /\partial
z$ which physically represents a curvature of the field lines over the thickness of the
sample. This bending is responsible for the increase of the rocking curve width
\cite{schelten}. This gives a direct measurement of the amount of transport current
flowing \textit{in the bulk} of the sample, via a simple integration of the
Maxwell-Ampere equation ($I_{bulk} \approx \frac{2W B}{\mu _o \Delta\omega}$). Bulk and
surface currents can thus be distinguished (see \cite{nous} for details). The
corresponding variation of $I_{bulk}$ as function of the applied current I is shown in
the fig. 4. It is clear that the large error bars due to the experimental resolution
(given mainly by the mechanical precision during the sample rotation) combined to the
relatively small number of experimental points do not allow to determine precisely the
distribution of the current. It is nevertheless quite reasonable to estimate that no bulk
current is present for $I<I_c$ and that a bulk current, roughly $(I-I_c)$, is observed
for $I>I_c$. The observed linear V(I) curve \cite{nous} and the critical current values
and variations \cite{patrice} are complementary indications in favor of a surface pinning
mechanism in NbSe$_{2}$.

 Concerning SANS measurements coupled with transport experiments in
NbSe$_{2}$, let us compare with Yaron et al experiments \cite{yaron}, which purpose was
to measure the longitudinal correlation length characteristic of FLL order. Yaron
$\textit{et al}$ observed a narrowing of the rocking curve when $I > I_c$. They
attributed this effect to an improvement in the FLL order. On the contrary, we observe
here, what was previously observed in other superconductors \cite{nous,schelten}, that
the rocking curve broadens when the over critical current penetrates the bulk. As this is
a simple consequence of the Maxwell equations, it appears not clear to us why such effect
was not observed by Yaron $\textit{et al}$. A possible interpretation is that the
reported Rocking curves are performed in the direction perpendicular to the one reported
in the present experiment. In such case and as observed in Nb-Ta samples \cite{houston},
a very small narrowing of the rocking curve can be observed. It can reasonably be
attributed to an enhancement of the homogeneity of the FLL Bragg planes spacing (due to
the homogeneous bending in the perpendicular direction) rather than to a change in a
correlation length. In any case, if this length can have a clear definition in static, it
has to be taken with care for $\it{curved}$ moving flux lines.

The FLL in NbSe$_{2}$, formed after a ZFC, appears to be quite similar to the FLL in
conventional type II superconductor with a moderate critical current. More differences
are expected after a FC, because in this case the V(I) curve exhibits a very peculiar
shape. The samples we used for the SANS experiments are larger than those usually
employed for transport properties and we have to precise that we have measured V(I)
curves (Fig.5c) very similar to what was already studied in details by others
\cite{capacitor,Ypaltiel,xiao}. In very short, they are hysteretic, with a S shape for
the first ramp of current after the FC and a linear shape and a reversible behavior for
all the following ramps.

Obtaining information on the FLL structure prepared after FC was not immediate. For the
same centering as for the ZFC case, any scattered intensity can be observed on the 2D
multi-detector. The first though was that the FLL was so strongly disordered that the
Bragg peaks were considerably broadened. But this is not the right reason, as evidenced
in Fig.5a where the corresponding $\omega$ rocking curve is shown. Compared with Fig. 3b,
one can realize that the Bragg condition has changed and above all that the rocking curve
exhibits a double peak, what is far from being expected. The sum of the integrated
intensity contained in these two peaks fits, within error bars, to the integrated
intensity of the Bragg peak of the FLL after ZFC, and the widths of the peaks are
comparable too. Consequently, we can not explain these strange Bragg peaks involving a
FLL disorder in its proper sense. More likely, they should correspond to two very similar
FLL which are ordered, but slightly tilted from the applied magnetic field direction from
few tens of degrees. An interpretation in terms of a rotation due to a Doppler shift can
be eliminated because the FLL frame is not moving. A more reasonable possibility is that
we are observing two FLL possessing two different Bragg planes spacing because of
different magnetic densities. This assumption would imply in the sample a magnetic field
gradient of more than half ($\Delta$B $\approx$ 0.2T) the applied one. This looks hardly
compatible with the strong interaction between the flux lines which limits the
compressibility of the vortex array. Furthermore, as evidenced in the figure 6, no change
is observed on the length of the Q vector, meaning that the average magnetic field
density inside the sample does not suffer from such strong heterogeneity.
 Finally, we propose that these two peaks are a signature of the "two phases" observed by Marchevski $\textit{et al}$ using
scanning hall ac probe \cite{twophases}. Their experiments evidenced that two states
possessing different critical currents are coexisting in the region of the peak effect.
Our SANS experiment offers complementary information. The fact that the two peaks are
very similar is not in favor of two states with a different bulk underlying disorder. The
shift between these two peaks indicates that the two FLL are slightly tilted by static
and small in plane field components. Let us call $+b_{1}$ and $-b_{2}$ these components.
The center of the Bragg peaks are turned by 0.13 and -0.40 deg from the initial Bragg
condition. It follows that $b_{1}$= 4000 tan(0.13) $\approx$ 9 G and $-b_{2}$ = 4000
tan(-0.35) $\approx$ -24 G. These field components should be induced by a peculiar and
non symmetric distribution of superficial currents. Following previous authors
\cite{Ypaltiel}, we adopt the point of view that the edge of the sample is the region of
the highest currents. Both Bragg peaks cover roughly the same surface. We can thus
estimate that the width of the sample is divided into two parts of the same size, i.e. 3
mm for each. We know that the low critical current is 2.5 A and that it corresponds to a
superficial value of $i_{low}= I_c/2W\approx 2 A/cm$. At the same time, we have measured,
when the peak effect is at its maximum, a ratio $\frac{I_{high}}{I_{low}}\approx 7$. With
the reasonable assumption that it corresponds to a state where the high critical current
state invades all of the sample, we can deduced that $i_{high}\approx 7 i_{low}\approx 14
A/cm$. Using the Ampere theorem and after a superposition with the top and bottom
surfaces, we find that the two domains transporting $i_{high}$ and $i_{low}$ generates
bulk components of magnetic field which are $b= \mu_o i\approx 2.5$ and $18$ G (Fig. 7).
This is not so far from the measured values (9 and 24 G), considering this highly
schematic picture.

In this picture, the FC state is characterized by large loops of current, that are as
many non dissipative paths for a transport current. Increasing the transport current
induces a preferential direction and one loop should be turned off. This implies that one
of the two tilted FFL has to disappear. This is indeed observed in fig.5b. If the
transport current is
 increased again up to the high critical current, the second loop disappear. All the flux lines
are now along the main magnetic field direction and the Bragg angle returns close to its
normal value (Fig.5c). Finally, the surfaces can not transport anymore non-dissipative
current, the excess penetrates the bulk and the current flow becomes resistive. The V(I)
curve returns to a classical behavior in the linear form $V=R_{ff}(I-I_{c})$, $I_c$ being
the low critical current. Since the initial loops have been cleaned by the current, they
have no reason to reappear and the V(I) curve is then observed reversible. We note that
this annealing-like effect by subcritical superficial current is in good agreement with
the magneto-optical results observed in the reference \cite{capacitor}.

 One of the remaining (and central) question is why FC is responsible for such a non usual state, whereas it
is classically the procedure used to obtain a FLL close to its equilibrium state. First,
this is correlated to the presence of the peak effect in the critical current. This
latter can be understood as superimposed on the "normal" critical current \cite{patrice}
\cite{hart}. If the sample is doped (here with Fe) or if impurities are present, the peak
effect generally broadens and the metastability observed in transport measurements after
FC become very obvious. This metastability is likely due to the (dynamical) coexistence
of large regions possessing two critical currents \cite{capacitor}, whom origin remains
unknown. We propose that these currents are in fact superficial. This has to be put close
to old results obtained in conventional type II superconductors: A thin surface film of
copper was shown to suppress the peak effect in Pb-Tl ribbons \cite{hart}, what evidences
a mechanism governed by surface currents. For a sample doped with impurities, small
broadening or even small differences between the critical fields, bulk $Bc_{2}$ or
surface $Bc_{3}$, can reasonably be expected. We can speculate that this is a reason for
a heterogeneous FC and for the corresponding freezing of metastable surface currents.
Looking at the influence of surface treatments such as metal plating in $NbSe_{2}$
appears thus to be particularly interesting in order to confirm the role of the surface
currents.

 In summary, we have studied by SANS the structure of the flux lines lattice and its link with dynamical properties. In Pb-In, a conventional type II superconductor,
V(I) curves are the same whatever the FLL state. On the contrary, a very peculiar
behavior is observed in Fe doped NbSe$_2$ for the FC case. The diffraction pattern
exhibits two FLL, shifted by tenth of degrees, corresponding to about tens of Gauss
perpendicular to the applied magnetic field. These extra-field components require the
presence of two loops of quasi-equilibrium currents possessing different values. This
leads to an annealing mechanism by sub critical currents confirming previous results
obtained by ac scanning probe \cite{capacitor}. Nevertheless, the FLL with the high
critical current appears not to be a bulk disordered state, like an amorphous or a glassy
state with a large amount of bulk free dislocations, but rather to be a state similar to
the ordered FLL with the moderate critical current. This can be explained by the
superficial nature of the critical currents. Of course, all questions concerning the
physical origin of this higher and unstable current remain open and more experiments are
needed to support this proposition.

Acknowledgments: It is a pleasure to thank Ted Forgan, Paul Kealey and Demetris
Charalambous (Birmingham University), Steve Lee (Saint Andrew) and the ILL staff (Bob
Cubitt and Charles Dewhurst) for their collaboration during the first SANS experiments on
Pb-In samples.
 J. Scola acknowledges support from "la region Basse Normandie".

\newpage

\begin{figure*}[tpb]
\centering \includegraphics*[width=15cm]{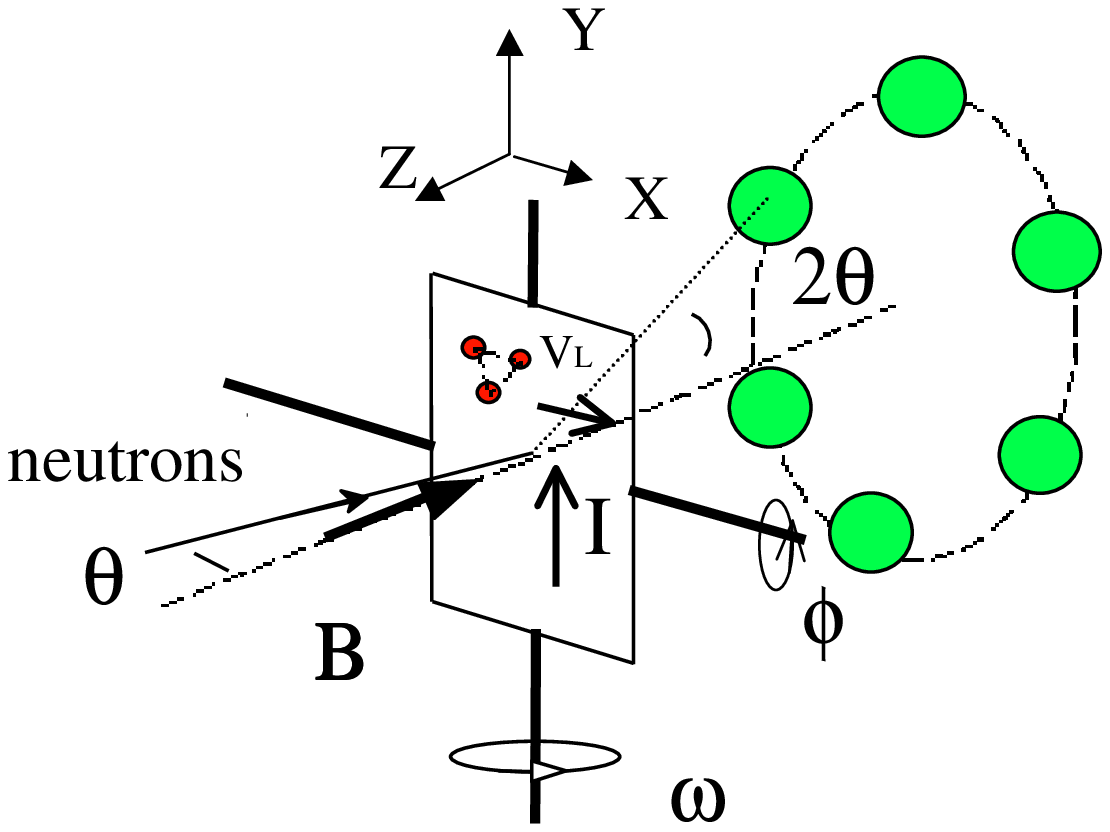} \caption{The geometry of the
experiment: The magnetic field $\vec{B}$ is applied parallel to the neutron beam and is
perpendicular to the large faces of the sample. The current $I$ flows vertically and for
$I > I_c$, vortex lines are moving perpendicular to it with a velocity $V_L$.}
\label{f.1}
\end{figure*}
\newpage
\begin{figure*}[tpb]
\centering\includegraphics*[width=14cm]{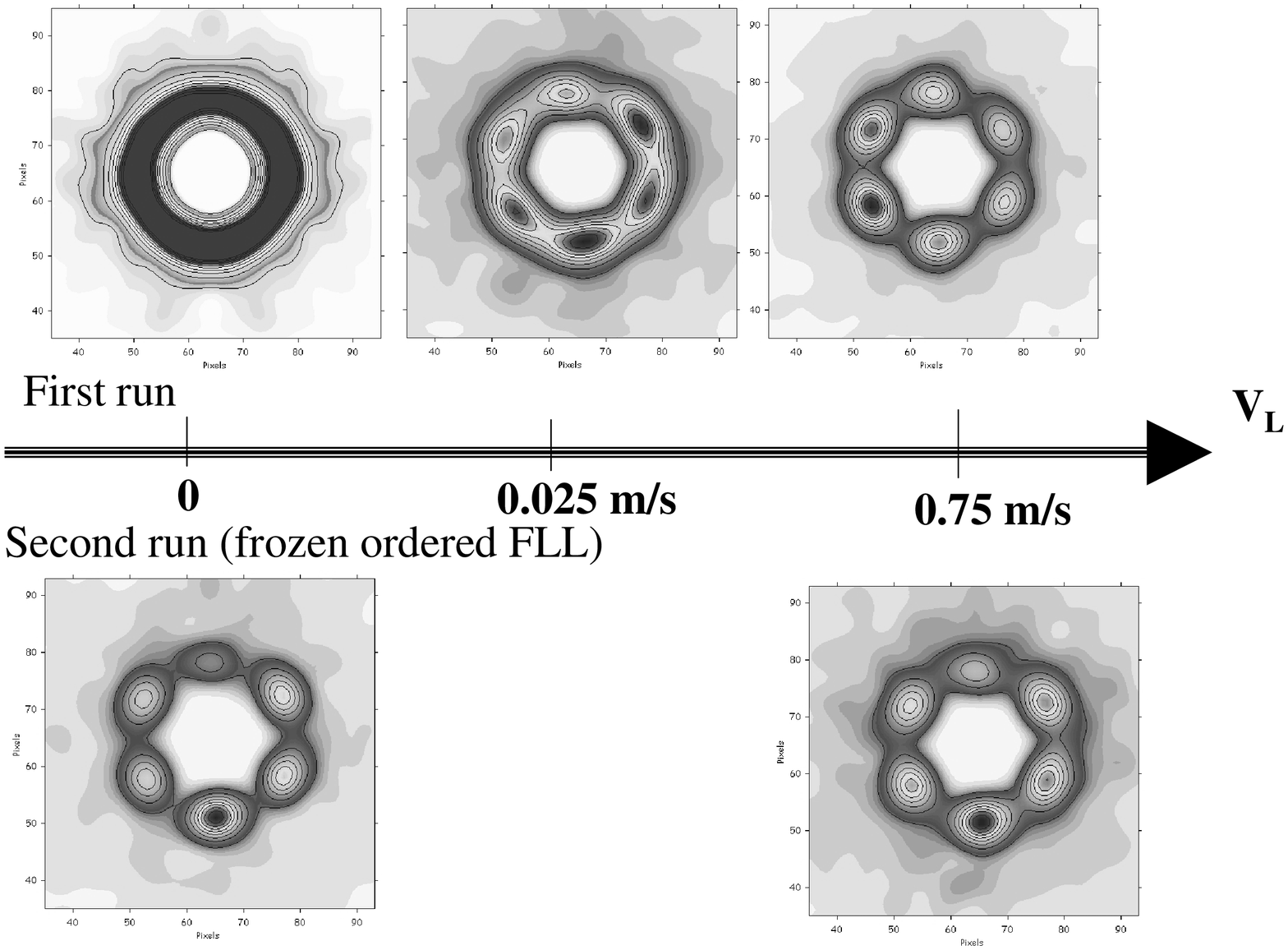}
\end{figure*}
\newpage
\begin{figure*}[tpb]
\centering\includegraphics*[width=13cm]{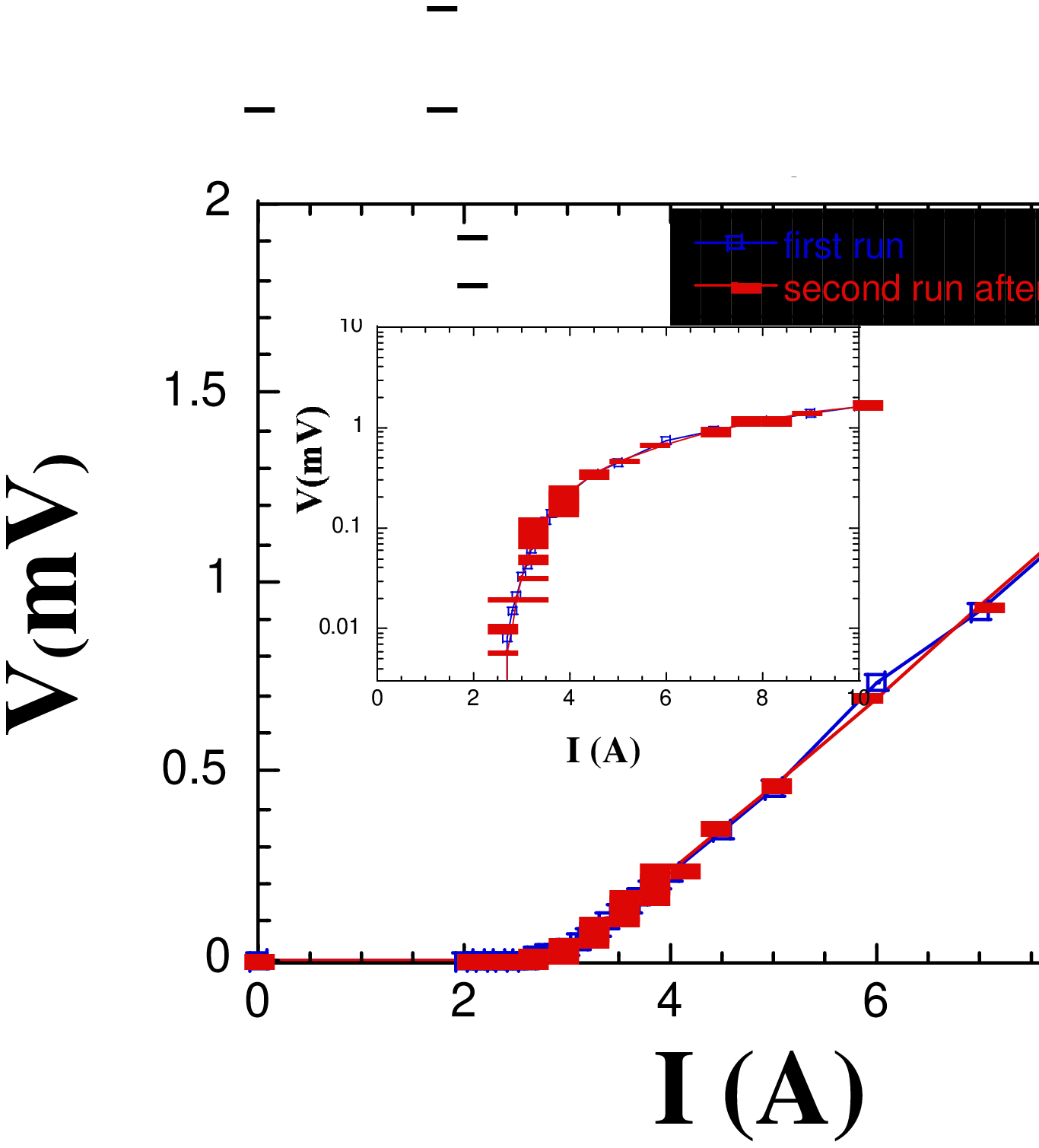}
 \caption{a/ 2D patterns of the FLL in a
polycrystalline Pb-In slab (T=2 K, B= 0.2 T). Top: For different velocities after the
equilibrium disordered state. Bottom: For different velocities after the frozen ordered
state. During the record, the sample is rocked as described in the text. b/ V(I) curve
for the two different initial states of the FLL (square:disordered and point:ordered). In
the inset is shown the same curve in a log-linear scale, so as to emphasize the perfect
similarity of the curves.}\label{f.2}
\end{figure*}
\newpage

\begin{figure*}
\includegraphics[width=18cm]{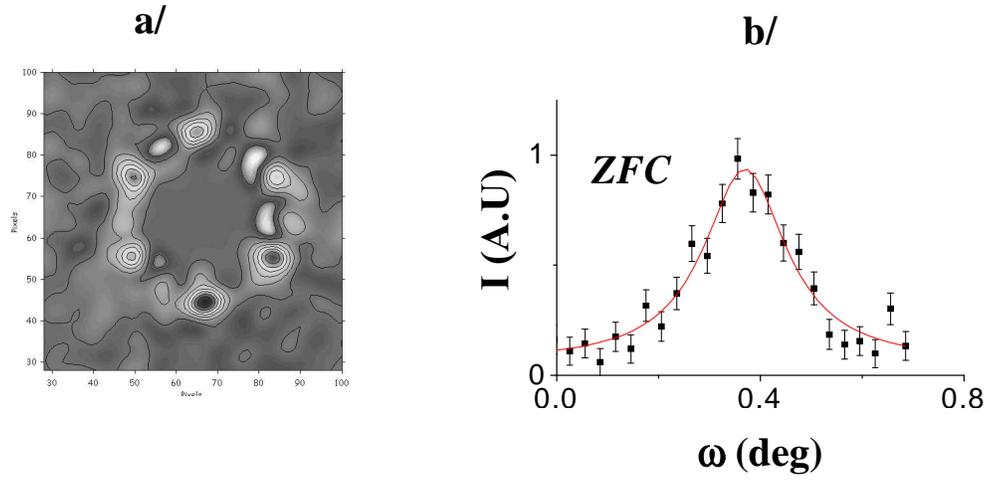} \caption{ a/ The diffraction pattern obtained on
the multi-detector at $2 K$, $0.4 T$ (ZFC) for the FLL in Fe doped NbSe$_{2}$. b/ The
corresponding  $\omega$-rocking curve. The fit is a Lorentzian ($\Delta\omega$ = 0. 234
$\pm$ 0.020 deg).}\label{f.3}
\end{figure*}
\newpage

\begin{figure*}
\includegraphics[width=13cm]{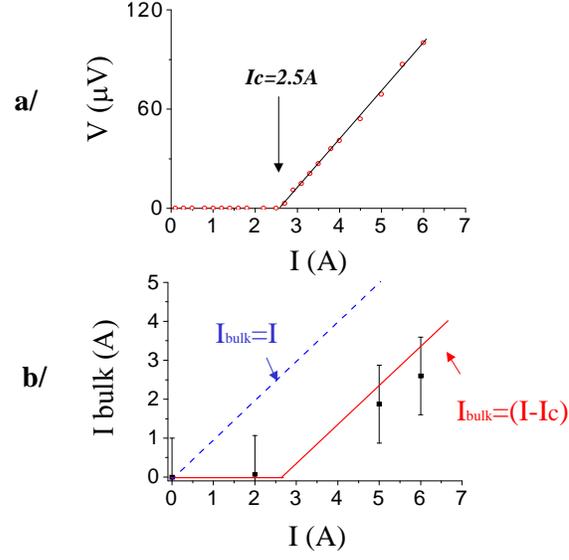}
\caption{a/ V(I) curve after a ZFC (2K, 0.4T). One can note the linear shape $V = R_{ff}
(I-I_{c})$. b/ The Bulk current versus the transport current flowing through the sample,
deduced from the broadening of the rocking curves (see text). The dotted line corresponds
to a homogeneous bulk current $I$. The solid line corresponds to a superficial current
$I_c$ and to a bulk over critical current $(I-I_c)$.}\label{f.4}
\end{figure*}
\newpage
\begin{figure*}
\includegraphics[width=15cm]{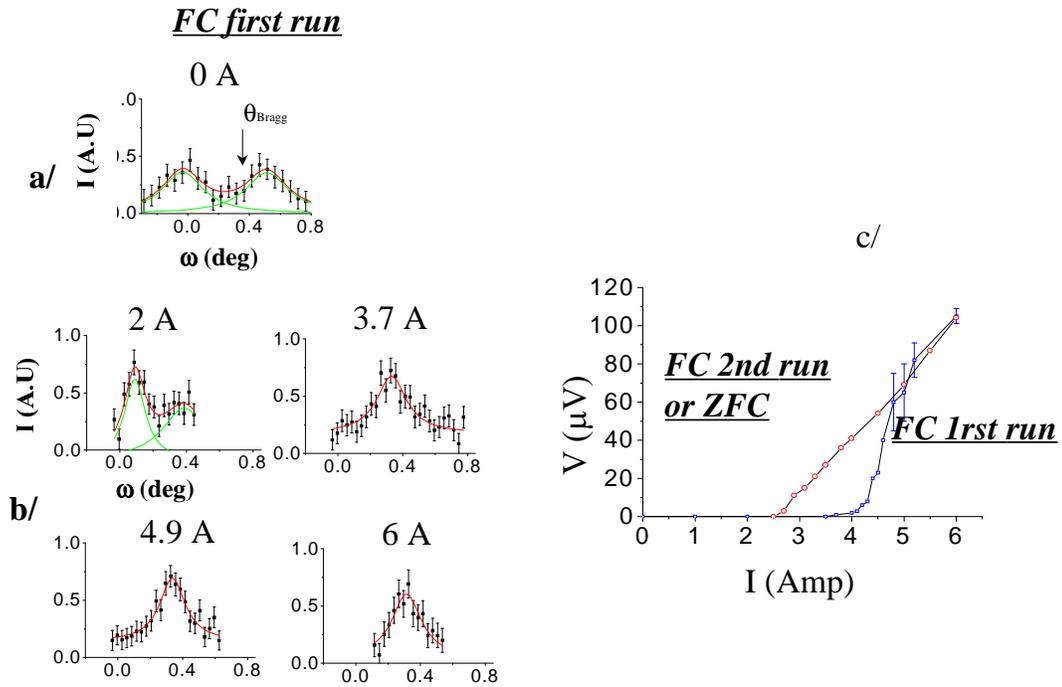}
\caption{a/ Rocking curve around $\omega$ for a Bragg peak (top right) of the FC FLL .
There is no applied current. The rocking curve fits with two Lorentzians.
       b/ The rocking curves for different values of the applied current after a FC. The high critical
       current is about 4 $\pm$ 0.5 A. Note the disappearance of one Bragg peak for a subcritical
       current of 2 A. At high current, the usual shape of the rocking curve is recovered.
       c/ The corresponding V(I) curve after a FC (S shape), and after ZFC or for the second ramp after FC (linear shape) (2K, 0.4T).}\label{f.5}
\end{figure*}
\newpage
\begin{figure*}
\includegraphics[width=20cm]{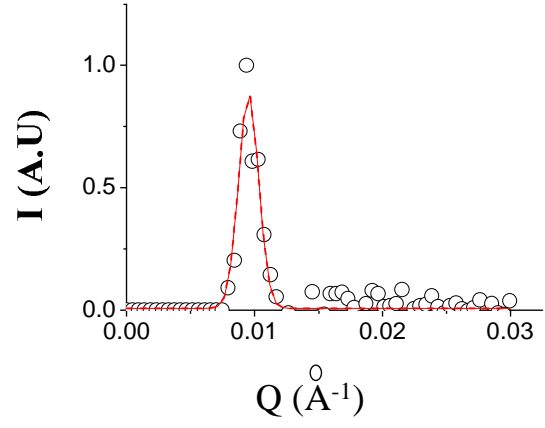}
\caption{The variation of the intensity diffracted from the FLL as function of the
diffraction vector Q (FC, B= 0.4T). The length of the Q vector corresponds to the value
fixed by the applied magnetic field of 0.4T. No obvious magnetic field gradients can be
observed.}\label{f.6}
\end{figure*}
\newpage
\begin{figure*}
\includegraphics[width=20cm]{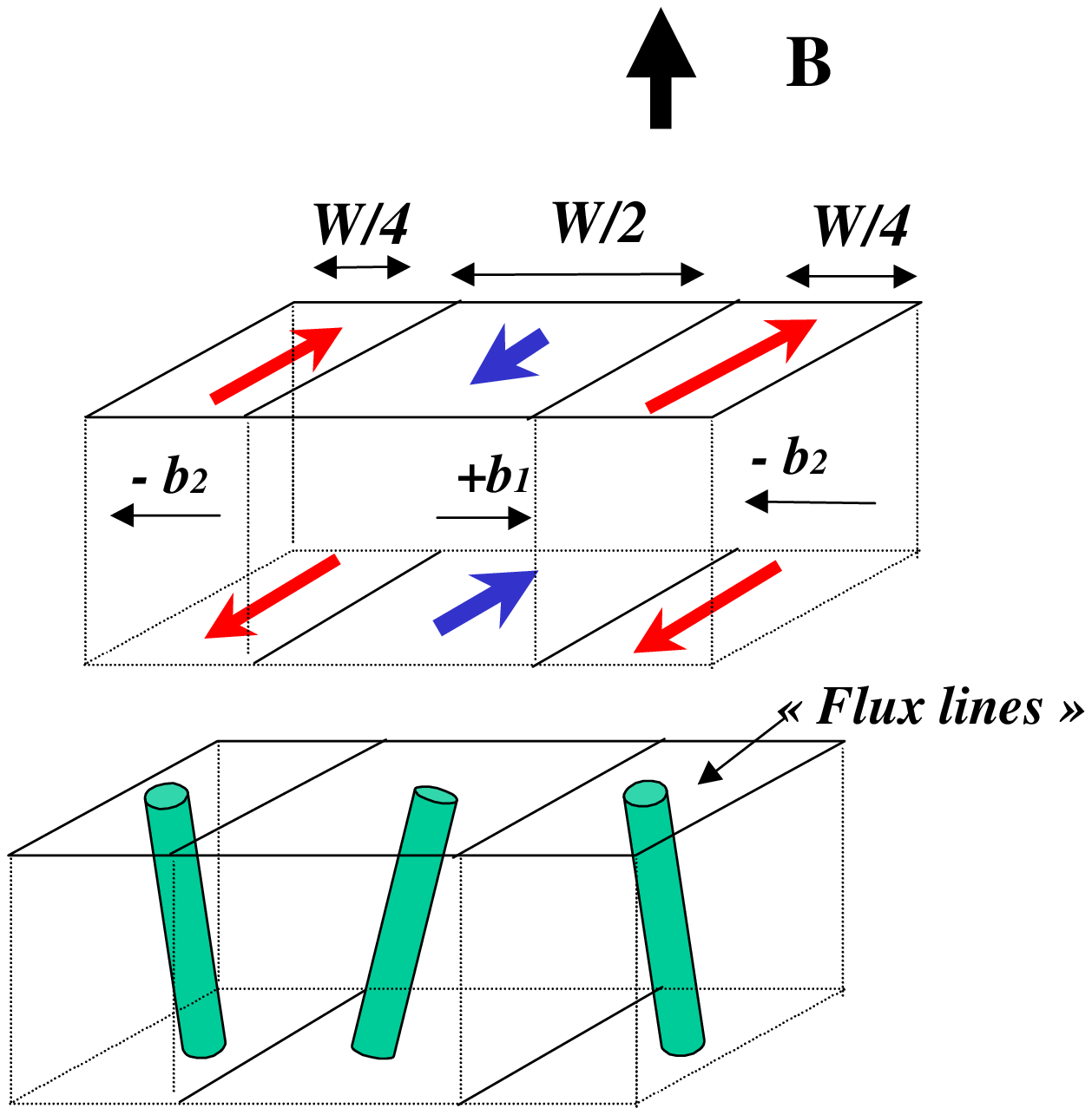}
\caption{One possible and highly schematic drawing of the current distribution in the
sample (without applied current). Only the top and bottom surfaces have been represented.
Small magnetic field components $+b_{1}$ and $-b_{2}$ of few Gauss are present in the
plane perpendicular of the applied magnetic field. This leads to two families of tilted
Flux Lines (bottom) responsible for the two Bragg peaks. For clarity, the shielding
diamagnetic currents are not represented.}\label{f.7}
\end{figure*}


\begin{references}


\bibitem{giam} T. Giamarchi , S. Bhattacharya, Vortex Phases, in ``High Magnetic Fields: Applications in Condensed Matter Physics and
Spectroscopy", p. 314, ed. C. Berthier et al., Springer-Verlag, 2002.

\bibitem{pippard} A.B. Pippard, Phil. Mag. 19, 217 (1969).

\bibitem{larkin} A.I. Larkin, Sov. Phys. JETP 31, 784 (1970). A.I. Larkin and Yu.N.
Ovchinnikov, J. Low Temp. Phys. 34, 409 (1979).

\bibitem{thorel} P.Thorel, Phd Thesis, University of Paris Orsay (1972).
Y. Simon and P. Thorel, Phys. Lett. 35A,
450 (1971).

\bibitem{giamar}T. Giamarchi and P. Le Doussal, Phys. Rev. Lett. 75, 3372 (1995).

\bibitem{ms} P. Mathieu and Y. Simon, Europhys. Lett. 5, 67 (1988).

\bibitem{archie} A.M. Campbell and J.E. Evetts, page  243 of "flux vortices and
transport currents in type II superconductors", Advances in physics  21, 199 (1972)

\bibitem{degennes} G. De Gennes and J. Matricon, Rev. Mod. Phys. 36, 45-49 (1964).

\bibitem{Kinnon} J.B. McKinnon, C.C. Chang and A.C. Rose-Innes, Proceedings of the
eleventh international conference on low temperature physics, Saint Andrews 1968, Edited
by J.F. Allen, D.M. Finlayson and D.M. McCall, 904 (1968).

\bibitem{Bhattacharya} S. Bhattacharya and M. J. Higgins, Phys. Rev. Lett. 70, 2620
(1993).

\bibitem{capacitor} M. Marchevsky, M. J. Higgins, and S. Bhattacharya, Phys. Rev. Lett. 88, 087002
(2002).

\bibitem{menghini} M. Menghini, Y. Fasano, and F. de la Cruz, Phys. Rev. B 65, 064510
(2002). Y. Fasano, M. Menghini, F. de la Cruz, Y. Paltiel, Y. Myasoedov, E. Zeldov, M. J.
Higgins, and S. Bhattacharya, Phys. Rev. B 66, 020512 (2002).

\bibitem{houston} C. Goupil, A. Pautrat, Ch. Simon, E.M. Forgan, P.G. Kealey, S.T. Johnson, G. Lazard,
 B. Pla\c{c}ais, Y. Simon, P. Mathieu, R. Cubitt, Ch. Dewhurst, Physica C 341, 999 (2000).
A. Pautrat, Ph.D. thesis, Universite de Caen (2000).



\bibitem{spinecho} E. M. Forgan, P. G. Kealey, S. T. Johnson, A. Pautrat, Ch. Simon, S. L. Lee, C. M. Aegerter, R. Cubitt, B. Farago, and P. Schleger
Phys. Rev. Lett. 85, 3488 (2000).

\bibitem{schelten} J. Schelten, H. Ullmaier, and G. Lippmann, Phys. Rev. B 12, 1772-1777 (1975)

\bibitem{nous} A. Pautrat, C. Goupil, Ch. Simon, D. Charalambous, E. M. Forgan, G. Lazard, P. Mathieu, and A.
Br\^{u}let, Phys. Rev. Lett. 90, 087002 (2003)

\bibitem{patrice} G. Lazard, P. Mathieu, B. Pla\c{c}ais, J. Mosqueira, Y. Simon, C. Guilpin, and G. Vacquier
Phys. Rev. B 65, 064518 (2002)

\bibitem{yaron} U. Yaron, P.L. Gammel, D.A. Huse, R.N. Kleiman, C.S.
Oglesby, E. Bucher, B. Batlogg, D. Bishop, K. Mortensen, K. Clausen, C.A. Bolle and F. De
La Cruz, Phys. Rev. Lett. 73, 2748 (1994). U. Yaron, P.L. Gammel, D.A. Huse, R.N.
Kleiman, C.S. Oglesby, E. Bucher, B. Batlogg, D. Bishop, K. Mortensen, K. Clausen, Nature
376 , 753 (1995).


\bibitem{Ypaltiel} Y. Paltiel, E. Zeldov, Y. Myasoedov, M. L. Rappaport, G. Jung, S.
Bhattacharya, M. J. Higgins, Z. L. Xiao, E. Y. Andrei, P. L. Gammel, and D. J. Bishop,
Phys. Rev. Lett. 85, 3712 (2000)

\bibitem{xiao} Z. L. Xiao, E. Y. Andrei, and M. J. Higgins, Phys. Rev. Lett. 83, 1664 (1999)

\bibitem{twophases} M. Marchevsky, M.J. Higgins and S. Bhattacharya, Nature (London)409,
591 (2001).

\bibitem{paltiel} Y. Paltiel, G. Jung, Y. Myasoedov, M. L. Rappaport, E. Zeldov, M. J. Higgins and S. Bhattacharya, Europhys. Lett., 58(1),112 (2002)

\bibitem{hart} P.S. Swartz and H.R. Hart Jr, Phys. Rev. 137, A818 (1965).


\end{references}
\end{document}